\documentclass{appolb}
\usepackage{epsfig}

\begin{document}
\title{Properties and Stability of Hybrid Stars%
\thanks{Presented at SQM 2011 in Krak\'ow}%
}
\author{S. Schramm, R. Negreiros, J. Steinheimer, T. Sch\"urhoff,
\address{FIAS, Ruth-Moufang-Str. 1, D-60438 Frankfurt am Main, Germany}
\and
V. Dexheimer
\address{Physics Department, Gettysburg College, Gettysburg, Pennsylvania, 17325 USA
}
}
\maketitle
\begin{abstract}
We discuss the properties of neutron stars and their modifications due to the occurrence of hyperons and quarks in the core of the star.
More specifically, we consider the general problem of exotic particles inside compact stars in light of the observed two-solar mass pulsar.
In addition, we investigate neutron star cooling and a possible explanation of the recently measured cooling curve of the neutron star in the supernova remnant Cas A.
\end{abstract}
  
\section{Introduction}

The study of strong interaction physics under extreme conditions is at the center of many experimental and theoretical efforts in nuclear physics.
Ultrarelativistic heavy-ion collisions allow for the investigation of the phase structure of QCD at high temperatures.  
Here one assumes that the conditions in the fireball of the collision zone generate chirally restored and deconfined matter, whose properties
can be deduced by analyzing observables like particle multiplicities and the collective flow.
At the other end of the QCD phase diagram, the investigation of neutron stars is the main tool to understand extremely dense and  cold matter.
As the theoretical approach of using lattice gauge simulations is not applicable in the high density/chemical potential regime one has to rely on model 
descriptions of the dense hadronic and possibly quark matter.
In the following we study such a model that includes hadronic flavor SU(3)  particles, as well as quarks, in an extension
of this approach. 

\section{Some Remarks on Observations and the Status of Theory}

A number of interesting observations in 2010 introduced new challenges to the theoretical understanding of compact stars.
Here, the accurate determination of the mass of pulsar PSR J1614-2230 of $M = 1.97 \pm 0.04 M_\odot$ \cite{197} is the
most prominent one. Meanwhile this value has been established as new benchmark for compact star modeling.
Another important measurement is the first observation of the real-time cooling behavior of a neutron star in the supernova remnant Cassiopeia A, 
where a rather steep drop of the surface temperature in the last 10 years has been recorded \cite{arXiv:1007.4719}. This result has significant impact
on cooling studies of compact stars.

For theory, especially the high value of the star mass implies the exclusion of a number of models. 
In general, but not necessarily, the more exotic the structure of the star the lower its mass becomes, given the simple picture that opening up new
degrees of freedom in dense matter will lead to a softening of the equation of state and consequently to smaller masses.
Thus, whereas a number of purely nucleonic equation of states yield large stellar masses, 
many calculations that include hyperons show maximum star masses far below the observed value (see e.g. \cite{arXiv:1006.5660,arXiv:1107.2497}).
Variation of the basic couplings of the hyperons can change this, as will be discussed further below and is it was also recently analyzed in \cite{arXiv:1112.0234}.


The situation is related but not quite the same in the case of hybrid stars that include a quark core.
In simple approaches that model the quark phase in terms of a bag model the mass of hybrid stars tends to be reduced significantly compared to the purely baryonic case
(see, for instance Fig. 2 of Ref.  \cite{arXiv:1011.2233}).
Depending on the value of the bag pressure the stars might become immediately unstable when quarks appear in the star with result that no stable
hybrid star exists. 
However, a more involved description of the quark phase potentially changes this picture as was already pointed out by Alford et al. in \cite{nucl-th/0411016}.
In this paper the authors show that perturbative QCD corrections can lead to an equation of state of the quark phase that is very similar to the hadronic one (in the
relevant range of densities). Therefore with such an approach stars with masses similar to purely nucleonic ones can be obtained. This study was also extended
to a wider range of parameters in \cite{arXiv:1102.2869}.
Instead of using a bag model description of quark matter one can describe the quark phase in a constituent quark picture like a Schwinger-Dyson approach
or NJL models. Bonanno et al. \cite{arXiv:1108.0559} have shown that within such an approach one can obtain a two-solar mass star including hyperonic matter and
a quark core. One requisite for this and other calculations of this type is a rather strong vector interaction term in the quark phase. 
It is worth noting, however, that quark models with such a term have serious problems in reproducing the lattice data of the behavior of the pressure at small chemical potentials as it was discussed in \cite{arXiv:1005.1176}.

\section{Our Specific Model}

In our approach we intend to combine hadronic and quark degrees of freedom in a unified way instead of dealing with two separate models for each phase.
This has the advantage that one can describe not only first-order phase transitions between hadrons and quarks but also second-order and cross-over transitions.
As we know from lattice QCD results that at low chemical potential the transition is a cross-over, an approach like the one formulated here is required in order to study
hadronic and quark regimes over the whole range of temperatures and densities.

Our studies are based on a hadronic flavor-SU(3) model that includes the lowest SU(3) multiplets for baryons and mesons. A detailed discussion of this ansatz can
be found in \cite{Papazoglou:1997uw,Papazoglou:1998vr}.  
The baryon-meson interaction term is given by
\begin{equation}
L_{Int}=-\sum_i \bar{\psi_i}[\gamma_0(g_{i\omega}\omega+g_{i\phi}\phi+g_{i\rho}\tau_3\rho)+m_i^*]\psi_i ,
\end{equation}
summing over the baryons $i$. The term includes the interaction with the non-strange and strange vector mesons $\omega, \rho$ and $\phi$.
The effective baryon masses $m_i^*$  are defined by the expression
\begin{equation}
m_i^* = g_{i\sigma} \sigma + g_{i\zeta} \zeta + g_{i\delta} \delta + \delta m_i ~~,
\end{equation}
including couplings to the non-strange scalar isoscalar $\sigma$, isovector $\delta$ and strange fields $\zeta$ plus a small explicit mass term.
Spontaneous chiral symmetry breaking is generated by the meson self-interactions, yielding non-vanishing vacuum expectation values of the
scalar fields, and in consequence, vacuum masses for the baryons.
The SU(3)-invariant self-interaction reads
\begin{eqnarray}
&L_{Self}=-\frac{1}{2}(m_\omega^2\omega^2+m_\rho^2\rho^2+m_\phi^2\phi^2)
+g_4\left(\omega^4+\frac{\phi^4}{4}+3\omega^2\phi^2+\frac{4\omega^3\phi}{\sqrt{2}}+\frac{2\omega\phi^3}{\sqrt{2}}\right)\nonumber&\\&+k_0(\sigma^2+\zeta^2+\delta^2)+k_1(\sigma^2+\zeta^2+\delta^2)^2+k_2\left(\frac{\sigma^4}{2}+\frac{\delta^4}{2}
+3\sigma^2\delta^2+\zeta^4\right)\nonumber\\&
+k_3(\sigma^2-\delta^2)\zeta+k_4\ \ \ln{\frac{(\sigma^2-\delta^2)\zeta}{\sigma_0^2\zeta_0}}~.&
\end{eqnarray}
An explicit chiral symmetry breaking term 
generates masses for the pseudoscalar mesons. The values
for the parameters can be found in \cite{Dexheimer:2008ax}.
 
\begin{figure}[th]
\centerline{\includegraphics[width=6.4cm]{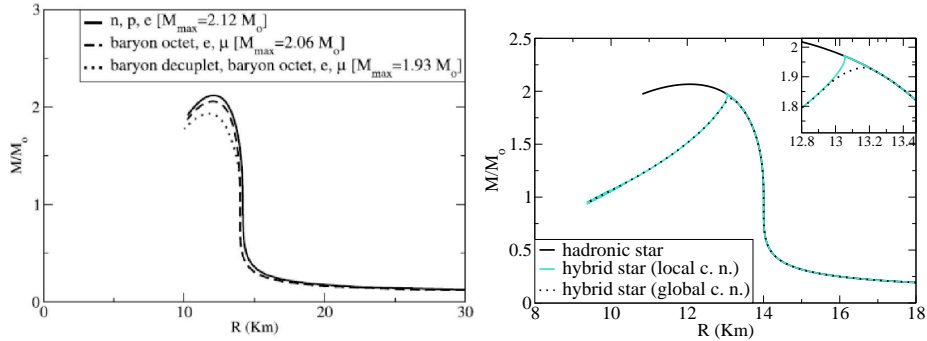}\includegraphics[width=5.8cm]{figs/Mass2.eps}}
\vspace*{5pt}
\caption{Left: Star masses as function of radii including nucleons, hyperons, and the baryonic spin 3/2 decuplet \protect\cite{Dexheimer:2008ax}.
Right: Mass-radius diagram of the quark-hadron model compared to the purely baryonic case. The inset shows the effect of introducing a Gibbs mixed phase.}
\label{mr}
\end{figure}

\begin{figure}[th]
\centerline{\includegraphics[width=5.4cm]{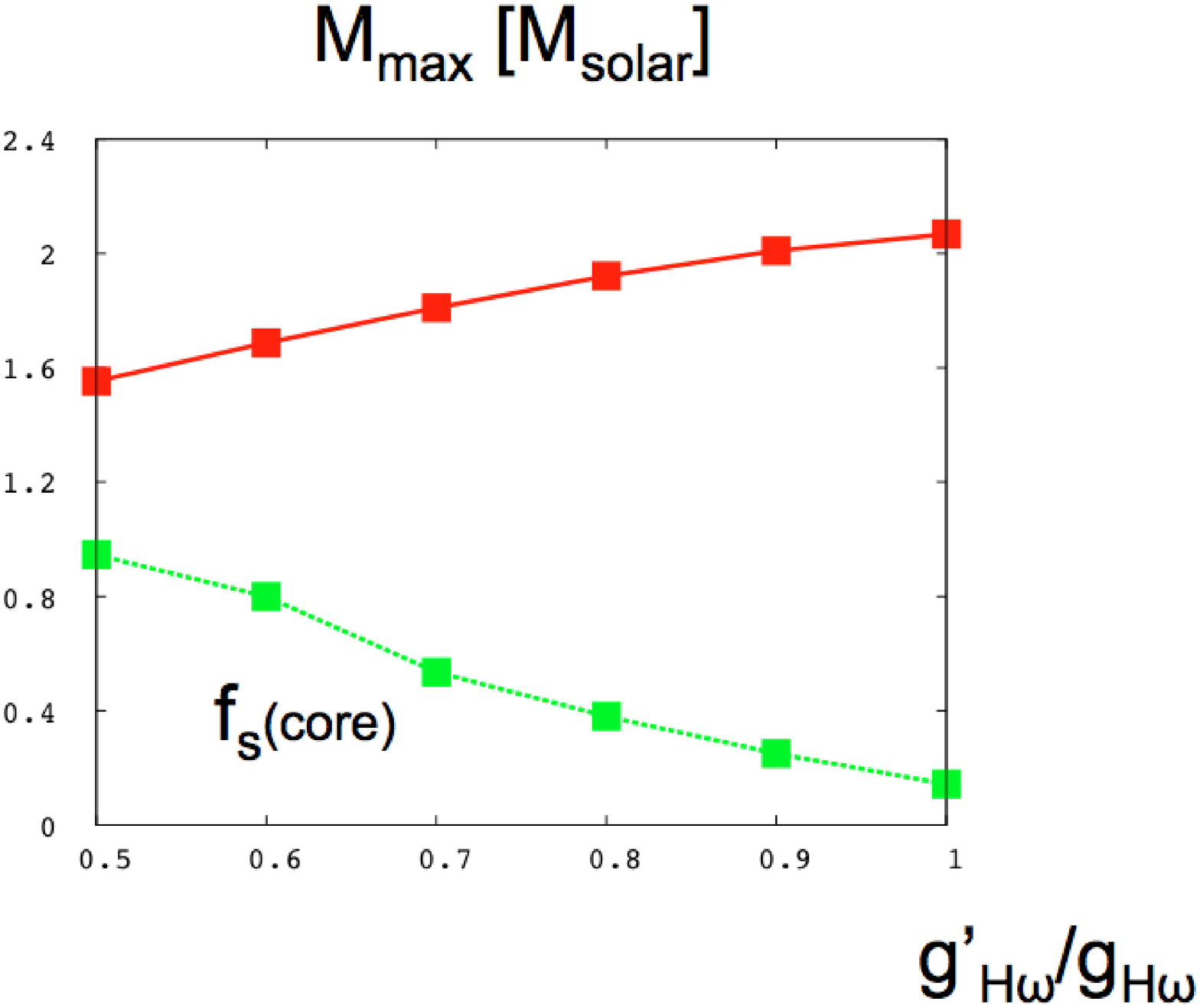}\includegraphics[width=6.9cm]{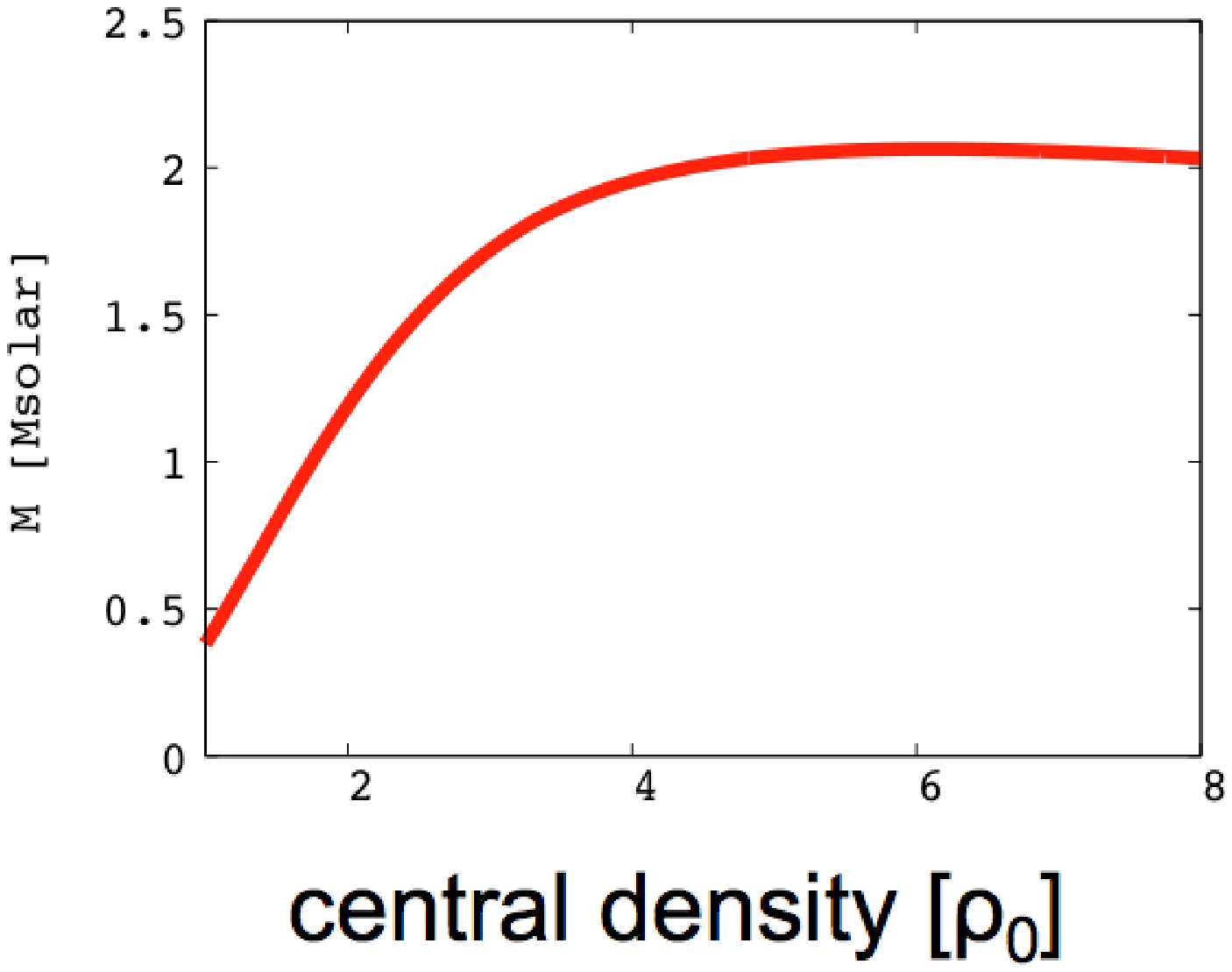}}
\vspace*{5pt}
\caption{Left: Maximum star mass and strangeness content $f_s$ of the star as function of the vector repulsion of the hyperons.
Right: Dependence of the baryonic star mass on the central energy density.}
\label{Mgh}
\end{figure}

Solving the Tolman-Oppenheimer-Volkoff equations for static spherical stars \cite{tov1,tov2} and including leptons for ensuring charge-neutral matter, one obtains masses and radii of stars as shown in Fig. \ref{mr}.
The results of three calculations are shown, for which different sets of degrees of freedom were taken into account.
Neglecting hyperons in the nucleonic case one obtains a maximum star mass of  2.12 solar masses. If one includes hyperons the
maximum mass drops slightly to 2.06 $M_\odot$. Further extending the possible baryonic degrees of freedom by including the spin 3/2 decuplet (mainly the
$\Delta$) leads to a mass of 1.93 $M_\odot$, which is still in agreement with observations. Especially in the latter case the amount of strangeness in the star is very
small as the $\Delta$ baryons essentially replace the $\Lambda$ and $\Sigma^-$ in the core of the star \cite{Dexheimer:2008ax}.
In contrast to the earlier discussion the hyperons do not have a significant impact on star masses. This is largely due to the meson interactions which keep the strange scalar field, compared to the non-strange field, rather big at high densities. In consequence the hyperons stay heavy and are not very strongly populated  \cite{Dexheimer:2008ax}.
In a schematic calculation one can observe the dependence of the amount of strangeness $f_s$ in the star and the maximum stellar mass.
Fig. \ref{Mgh} shows the result of a calculation using the same model but artificially reducing the vector coupling constant of the hyperons at densities beyond nuclear matter densities (thus without changing the reasonably well-known optical potential depths in normal nuclear matter). For instance, reducing this value by 50 percent increases the strangeness fraction in the core of the star from 0.1 to 1, that is, on the average one strange quark per baryon. At the same time the maximum mass is significantly reduced by half a solar mass. It might be interesting to collect similar results for the relative reduction of the star mass as function of strangeness content for  a range of models to see
whether this behavior is qualitatively and quantitatively similar over a wider range of theoretical approaches.

We include quark degrees of freedom in the model description in a similar way as it is done in the so-called PNJL approach \cite{Fukushima:2003fw,Ratti:2005jh}.
The quark fields couple to the scalar and vector condensates. As an effective field describing the deconfinement phase transition we
introduce the field $\Phi$, in analogy to the Polyakov loop field in the PNJL models.
This field couples to the hadron and quark
masses such that quarks attain a high mass in the confined phase at low values of $\Phi$ and correspondingly hadrons obtain a large mass for large values of the $\Phi$, removing the baryons as degrees of freedom  (see Ref. \cite{Dexheimer:2009hi}).
The potential $U(\Phi)$ of the field is chosen in a way to reproduce lattice data for energy density and pressure at zero chemical potential for a wide range of temperatures.
It reads
\begin{equation}
U = (a_0 T^4+a_1 \mu^4+a_2 T^2 \mu^2) \Phi^2 + a_3  T_0^4 \log{(1-6\Phi^2+8\Phi^3-3\Phi^4)}.
\end{equation}
The values of the parameters are quoted in \cite{Dexheimer:2009hi}.
In addition to the usual structure of the potential we add the chemical-potential dependent terms proportional to $a_1, a_2$ in order to reproduce
a critical end point of a first-order phase transiton line at larger values of the chemical potential 
as suggested by lattice calculations \cite{Fodor:2004nz}.
The resulting star masses and radii are shown in the right panel of Fig. \ref{mr}.
The onset of the quark phase essentially leads to an unstable system, which reduces the maximum mass of attainable neutron stars in this model.
However, the reduction of mass is a moderate drop of 10 percent with respect to the value without quarks, which is still in agreement with observation.
If one assumes global charge neutrality one obtains a Gibbs mixed phase of quarks and baryons in the inner 2 km core of the star.
This small change of the maximum mass in spite of the onset of a new exotic phase can be understood, looking at the right panel of Fig. \ref{Mgh}.
Here the mass of the baryonic star is plotted as function of central density. The maximum mass is reached at roughly 6 $\rho_0$. Including quarks in this model, the quarks
are populated at densities beyond $\rho \approx 4 \rho_0$. As an extreme assumption, this limits the stable star solutions to this maximum value of density. However, the graph
shows that the density dependence of the mass on the central density is pretty much flat over a wide range of densities. Therefore a reduction of the maximum central
density of even 40 percent by the occurrence of some exotic component like quarks or a kaon condensate 
has little impact on the maximum star mass.

\section{Rotating Stars}

\begin{figure}[th]
\centerline{\includegraphics[width=6.4cm]{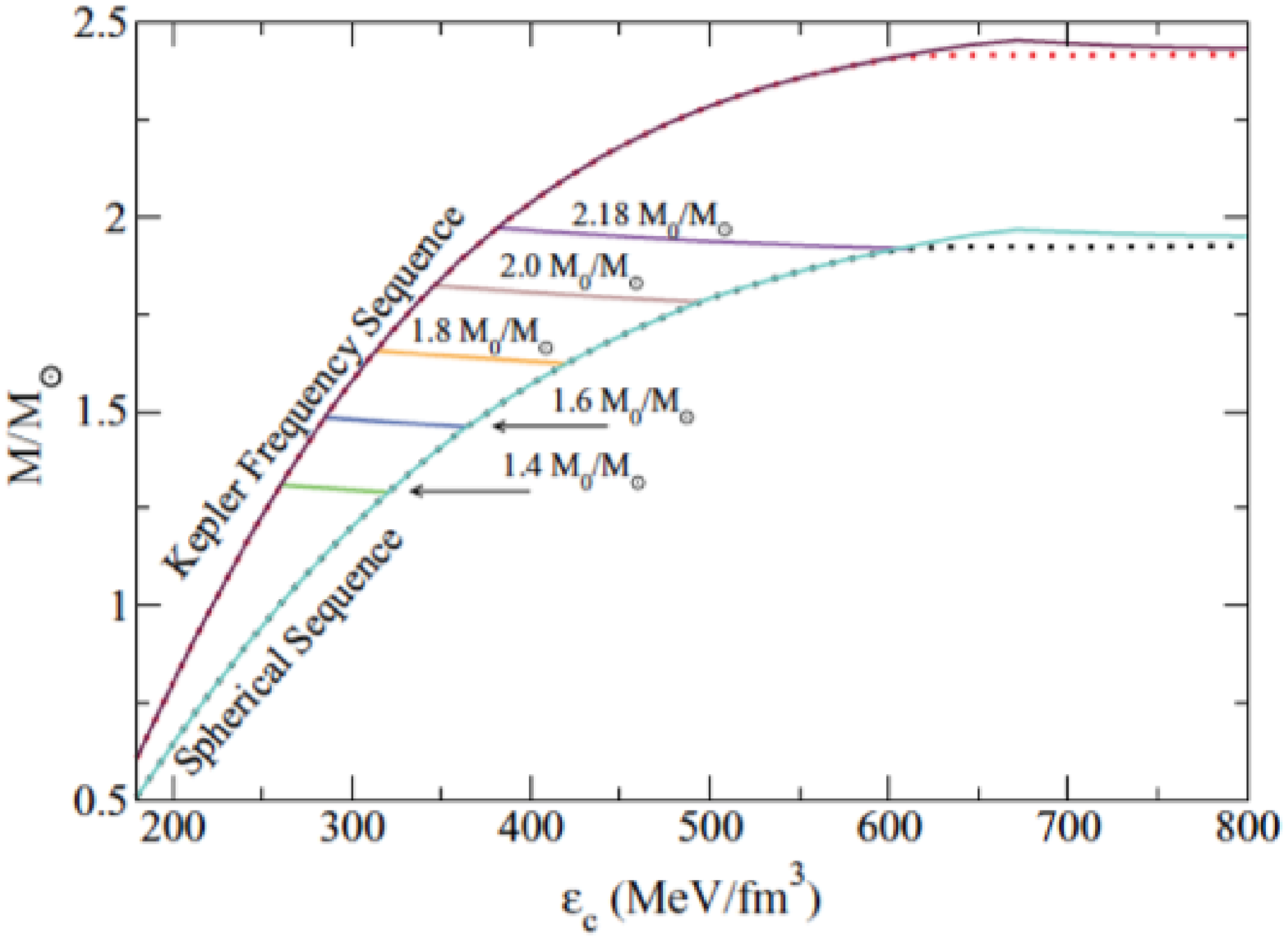}\includegraphics[width=6.4cm]{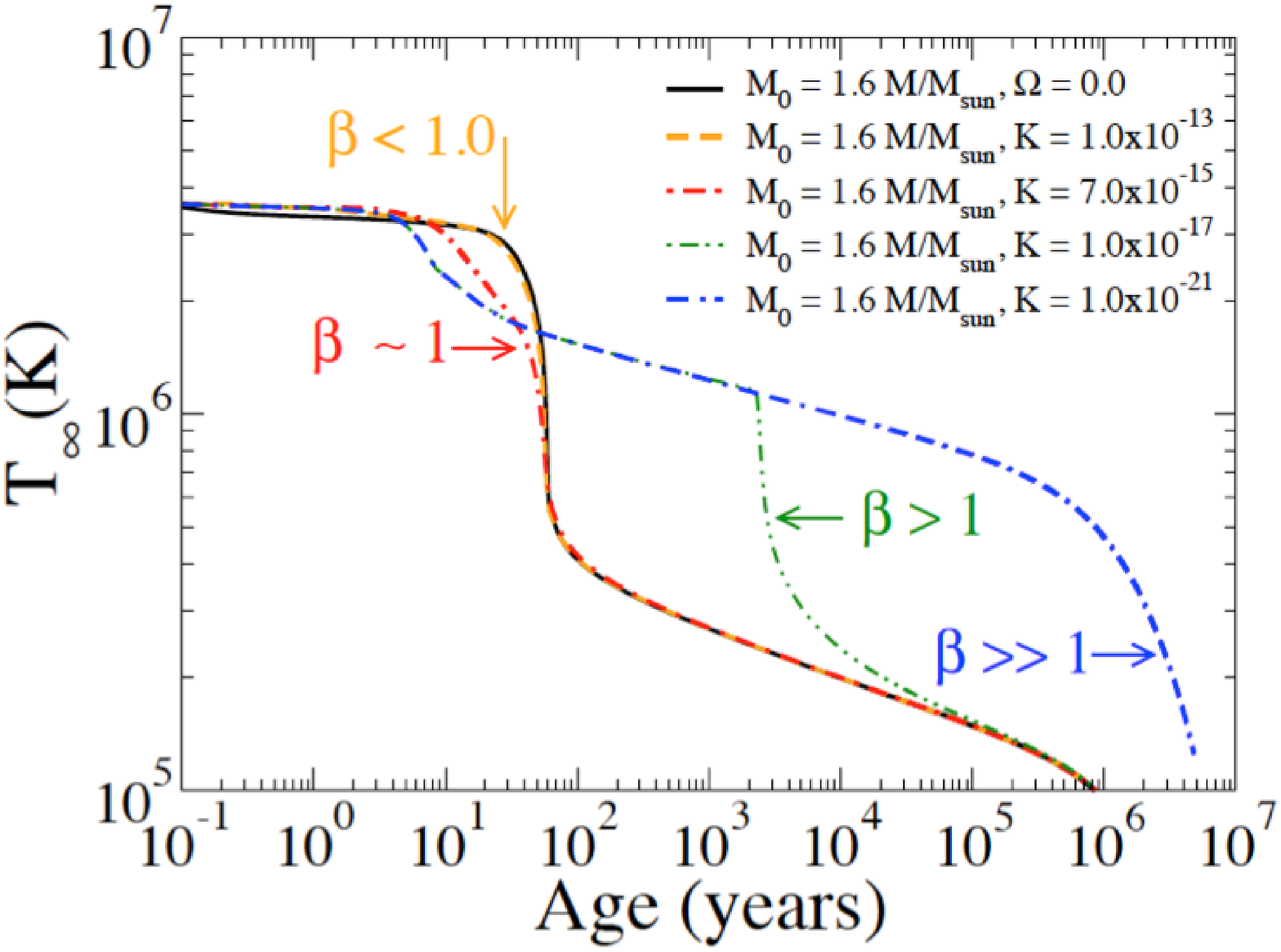}}
\vspace*{5pt}
\caption{Left: Star masses as function of central energy density. Results for the static solutions and for stars rotating at their Kepler frequency
are shown. The nearly horizontal lines show the gravitational mass spinning-down stars with a constant baryon number.
Right: Cooling curves of neutron stars depending on the ratio $\beta$ of characteristic time scales  (see text).}
\label{Afixed}
\end{figure}

Rotating stars can sustain a substantially larger gravitational mass than in the static case. Results for our model are shown in Fig. \ref{Afixed}.
Lines for the static solutions and for stars spinning at their Kepler frequencies are shown as function of the central energy density. One can observe
roughly a 20 percent increase in the maximum mass. However, if one considers the time evolution of a star without mass accretion or ejection, like in the 
case of an isolated neutron star or for a neutron star in a binary system after the accretion phase, the baryon number (or the corresponding so-called baryon
mass) stays constant during the slowing-down evolution of the star. The results for various fixed baryonic masses are shown as nearly 
horizontal lines  in the same figure, which shows that the gravitational mass stays nearly constant. 
On the other hand the central energy density changes substantially during the evolution 
by potentially more than 50 percent, drastically altering the particle populations inside of the star. In general this can lead to large changes in the star's cooling behavior. Looking at the $\beta < 1$ curve ($\beta$ is defined below) of Fig. \ref{Afixed} (right panel) one can observe the standard behavior of the cooling curve of a massive star.
The core of the star cools mainly by neutrino emission via the direct Urca processes like $n \rightarrow p + e^- + \bar{\nu}_e$. This channel only operates
at higher densities, as it requires a density ratio of  protons to neutrons of about $1 / 9$, so that energy and momentum are conserved in the process.
The cooling wave from the core propagates to the surface, which is reached after about 100 years leading to a sudden drop of the temperature.
However, in case the star is rotating, the results depend on the ratio $\beta$ of the spin-down time to the core-crust coupling time of 100 years
\cite{Negreiros:2011ak}. If the star rotates sufficiently fast, the central density of the star might be too low for the direct Urca process to take place,
therefore suppressing the fast cooling channel. During the spin-down of the star this threshold will be reached at a later time, when the direct Urca process
leads to a delayed strong drop in temperature as shown in the $\beta > 1$ curve. 
As mentioned before, for the first time the cooling evolution of a star has been observed, showing a strong drop in temperatures, albeit at a time of more than 300 
years after the supernova explosion. If this steep decline wer caused by the direct Urca process one would have expected it much earlier and the temperatures of the star would have
dropped to much lower, unobservable, values, than the ones seen. However, in case the star rotates, this situation can be easily explained as the delayed fast cooling
process, which was discussed above. The corresponding fit curves to the data for different star masses are shown in the figure. For a more detailed explanation of the parameters entering this analysis 
see \cite{Negreiros:2011ak}. Alternative explanations of the drop of temperature, which have to assume that no direct Urca process is present in the star, relate this drop
to the onset of neutron superfluidity\cite{arXiv:1012.0045,arXiv:1011.6142}.

\begin{figure}[th]
\centerline{\includegraphics[width=6cm]{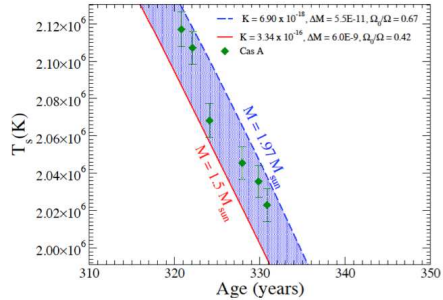}}
\vspace*{5pt}
\caption{Temperature data as measured for the Cas A neutron star. The lines show fits to the data assuming two different star masses. The shaded area
between the curves can be covered by parameter adjustments.}
\label{CasA}
\end{figure}

\end{document}